\documentclass[epjST]{svjour}
\usepackage{graphicx}
\usepackage{color}
\usepackage[percent]{overpic}
\usepackage{amsmath,amssymb,amsfonts}
\usepackage[normalem]{ulem}
\usepackage{verbatim}
\usepackage[colorlinks=true,linkcolor=blue,urlcolor=blue,citecolor=blue]{hyperref}

\newcommand {\mc} {\mathcal}
\newcommand {\lan} {\left \langle}
\newcommand {\ran} {\right \rangle}

\newcommand {\sign} {\mathrm{sign}}

\definecolor{Green}{rgb}{0,0.6,0}

\begin{document}
\title{Hong-Ou-Mandel characterization of multiply charged Levitons}

\author{Dario Ferraro\inst{1}\fnmsep\thanks{\email{ferraro@fisica.unige.it}} \and Flavio Ronetti\inst{2,3,4} \and Luca Vannucci\inst{2,3} \and Matteo Acciai\inst{2,3,4} \and J\'er\^ome Rech\inst{4} \and Thibaut Jockheere\inst{4} \and Thierry Martin\inst{4} \and Maura Sassetti\inst{2,3}}
\institute{Istituto Italiano di Tecnologia, Graphene Labs, Via Morego 30, I-16163 Genova, Italy \and 
Dipartimento di Fisica, Universit\`a di Genova, Via Dodecaneso 33, 16146, Genova, Italy \and 
CNR-SPIN, Via Dodecaneso 33, 16146, Genova, Italy \and
Aix Marseille Univ, Universit\'e de Toulon, CNRS, CPT, Marseille, France}
\abstract{
We review and develop recent results regarding Leviton excitations generated in topological states of matter - such as integer and fractional quantum Hall edge channels - and carrying a charge multiple of the electronic one. The peculiar features associated to these clean and robust emerging excitations can be detected through current correlation measurements. In particular, relevant information can be extracted from the noise signal in generalized Hong-Ou-Mandel experiments, where Levitons with different charges collide against each other at a quantum point contact. We describe this quantity both in the framework of the photo-assisted noise formalism and in terms of a very interesting and transparent picture based on wave-packet overlap.
}
\maketitle
\section{Introduction}
\label{intro}

The on-demand generation of single-electron states in mesoscopic systems has opened the way to the fascinating field of electron quantum optics (EQO), where individual fermionic excitations ballistically propagating in solid state systems are manipulated with methods borrowed from photonic quantum-optical experiments \cite{Grenier11,Bocquillon14,Bauerle18}. The standard recipe of EQO starts from the on-demand generation of coherent single-particle excitations in a solid state device, which can be achieved using the so-called mesoscopic capacitor \cite{Feve07,Moskalets08} or driving the quantum conductor with carefully-engineered periodic voltage pulses \cite{Dubois13,Dubois13b,Gabelli13,Dolcini16}. The dissipationless edge states of the quantum Hall effect are then exploited to ensure ballistic propagation of the wave-packet throughout the sample, and quantum point contacts (QPC) play the role of beam-splitter allowing to conceive efficient interferometric setups.

However, it is well known that solid state systems can be heavily influenced by interactions, differently from what happens in the traditional photonic quantum optics. For instance, the fractional quantum Hall (FQH) effect \cite{Tsui99} is a paradigmatic example of the dramatic consequences of electron-electron interactions. Here, a new strongly-correlated phase emerges in the quantum liquid, with quasiparticle excitations carrying a fraction of the electron charge \cite{Laughlin83,DePicciotto97,Saminadayar97} and whose statistical properties are neither bosonic nor fermionic, but belong to the more general class of anyons \cite{Stern08}. Since FQH systems host topologically protected chiral modes at the edge \cite{Wen95}, an extremely exciting generalization of EQO ideas to this strongly interacting regime can be envisaged.

With the present paper we intend to review and extend recent results concerning EQO in a quantum Hall fluid at integer and fractional filling factor. In particular, we focus on a solid state analogue of the celebrated Hong-Ou-Mandel (HOM) experiment \cite{Hong87}, where excitations impinge on the opposite sides of the beam splitter. In this experimental configuration it is possible to extract valuable information about the statistical properties of the colliding particles and the form of their wave-packets. The simplest way to generate clean pulses in the input arms of the HOM interferometer is by means of a periodic Lorentzian voltage drive, which was shown to provide on-demand robust multi-electronic excitations both in the integer \cite{Levitov96,Keeling06} and the fractional regime \cite{Rech17,Vannucci17}. According to existing literature, throughout this review we will refer to these excitations carrying $q$ elementary charges as $q$-Levitons \cite{Ronetti17}. We find that, in strongly-correlated FQH states, the HOM noise generated by the collision involving $q$-Levitons with $q>1$ shows unexpected sub-dips \cite{Ronetti17} which are consistent with an interaction-induced self-organization of the multi-electronic wave-packet in the time domain, namely a real-time version of the crystallization occurring in Luttinger liquids confined by a potential well \cite{Pecker13,Gambetta14}. This work collects evidence of the prominent role played by HOM collisional experiments as a tool to characterize the behavior of individual excitations ballistically propagating in condensed matter systems \cite{Ferraro17}.

This review is organized as follows. We illustrate the model for a Hall bar in the Laughlin sequence of the FQH effect in presence of the periodic drives in Section \ref{Model}. Then, in Section \ref{Tunneling}, we introduce a quantum point contact and evaluate the corresponding shot-noise generated by tunneling between the two edges. In Section \ref{Results} we apply such formalism to study HOM interferometry in EQO. We review recent theoretical and experimental results at integer and fractional filling factor, and discuss new unexpected features emerging from generalized HOM collisions in the FQH regime. 
Later, we also discuss the relation between current noise and the overlap of Levitons wavefunctions in Section \ref{wavepacket}.
Finally, in Section \ref{Conclusions}, we draw our conclusions.

\section{Model}
\label{Model}

\begin{figure}
	\centering
	\includegraphics[width=0.75\linewidth]{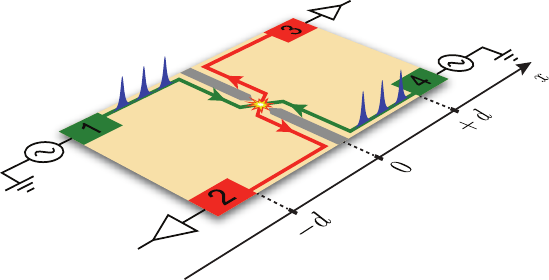}
	\caption{Four-terminal setup for Hong-Ou-Mandel interferometry in the FQH regime. Periodic trains of Levitons (blue peaks) are injected through input terminals 1 and 4, while contact 2 and 3 are the output terminals where the current and the noise are measured.}
	\label{fig:setup_hom}
\end{figure}

Let us consider a Hall bar in a four-terminal geometry, such as the one depicted in Figure\ \ref{fig:setup_hom}. For a filling factor $\nu$ in the Laughlin sequence $\nu=1/(2n+1)$ \cite{Laughlin83}, with $n \in \mathbb N$, a single chiral mode propagates at each edge of the sample. Right-moving and left-moving low-energy degrees of freedom are described by bosonic fields $\Phi_{R/L}$ satisfying $[\Phi_{R/L}(x), \Phi_{R/L}(y)] = \pm i \pi \sign(x-y)$. The effective Hamiltonian for the edge modes reads \cite{Wen95} ($\hbar=1$)
\begin{equation}
	H_{R/L} = \frac{v}{4\pi} \int dx \left[ \partial_x \Phi_{R/L}(x) \right]^2  ,
\end{equation}
with $v$ the propagation velocity of the free chiral fields.
We assume that the edge modes are capacitively coupled with two oscillating voltage drives $\mc V_R(x,t)$ and $\mc V_L(x,t)$, acting separately on the right-moving and left-moving excitations respectively. To this end, we introduce the coupling term
\begin{equation}
	H_V = e \int dx \mc V_R(x,t) \rho_R(x) + e \int dx \mc V_L(x,t) \rho_L(x) ,
\end{equation}
where the density operators $\rho_{R/L} (x) = \mp \frac{\sqrt{\nu}}{2\pi} \partial_x \Phi_{R/L}(x)$ have been introduced. To describe the experimentally relevant situation of infinite, homogeneous contacts we factorize the space and time dependence of the voltage sources as $\mc V_{R/L}(x,t) = \Theta(\mp x-d) V_{R/L}(t)$. Under such assumption, the equations of motion $(\partial_t \pm v \partial_x) \Phi_{R/L}(x,t) = e \sqrt \nu \mc V_{R/L}(x,t)$ are solved by
\begin{equation}
	\Phi_{R/L}(x,t) = \phi_{R/L}(x \mp vt,0) + e \sqrt \nu \int_0^{t \mp \frac x v - \frac d v} dt' V_{R/L}(t') .
\end{equation}
In the above equation, we have indicated with $\phi_{R/L}$ the free chiral bosonic fields at zero voltage (equilibrium configuration with $V_L=V_R=0$). The unimportant constant time shift $d/v$ will be omitted in the rest of the paper. Finally, the standard procedure of bosonization allows one to define annihilation fields for Laughlin quasiparticles as \cite{Wen95}
\begin{equation}
\label{eq:bosonization}
	\Psi_{R/L}(x,t) = \frac{F_{R/L}}{\sqrt{2\pi a}} e^{\pm i k_{\rm F} (x \mp vt)} e^{-i \sqrt \nu \phi_{R/L}(x \mp vt,0)} e^{-i e^* \int_0^{t \mp \frac x v} dt' V_{R/L}(t')} ,
\end{equation}
where $a$ is a short-distance cut-off, $k_{\rm F}$ is the Fermi momentum, $F_{R/L}$ are the Klein factors \cite{Guyon02,Martin05,Dolcetto13} and $e^*=\nu e$ ($e$ the electron charge) indicates the fractional charge of a quasiparticle in the Laughlin sequence \cite{Laughlin83}. 

\section{Shot-noise induced by tunneling}
\label{Tunneling}

A quantum point contact (QPC) placed at $x=0$ induces tunneling of Laughlin quasiparticles between opposite edges \cite{Kane92}. This is accounted for by the tunneling Hamiltonian
\begin{equation}
	H_{\rm tun} = \Lambda \Psi^\dag_R(0) \Psi_L(0) + \mathrm{h.c.} ,
\end{equation}
with $\Lambda$ the constant tunneling amplitude. 
Operators for the right and left moving currents are readily obtained from continuity equations, and read $J_{R/L}(x,t) = -v \frac{e \sqrt \nu}{2\pi} \partial_x \Phi_{R/L}(x,t)$ downstream of the input terminals. Considering the tunneling as a perturbation, we can write the time-dependent current operators up to second order in $H_{\rm tun}$ as
\begin{align}
	J_{R/L}(x,t)
	& = -v \frac{e \sqrt \nu}{2\pi} \partial_x \phi_{R/L}(x\mp vt,0)  \pm\frac{e^2\nu}{2\pi}V_{R/L}(t\mp\frac{x}{v}) + \nonumber \\
	& \quad - i \int_{-\infty}^t dt' \left[ H_{\rm tun}(t'), v \frac{e \sqrt \nu}{2\pi} \partial_x \phi_{R/L}(x \mp vt) \right] + \nonumber \\
	& \quad + \int_{-\infty}^t dt' \int_{-\infty}^{t'} dt'' \left[  H_{\rm tun}(t'') , \left[ H_{\rm tun}(t'), v \frac{e \sqrt \nu}{2\pi} \partial_x \phi_{R/L}(x \mp vt) \right] \right] + \nonumber \\
	& \quad + o(H_{\rm tun}^2).
\end{align}
Notice that, according to the analysis carried out in Ref. \cite{Crepieux04}, the present perturbative approach is physically meaningful as long as the backscattering conductance associated to the quasiparticle tunneling at the QPC is smaller than the quantized conductance of the quantum Hall channel.

We define the currents flowing into output terminals 2 and 3 as $J_3(t) = J_R(d,t)$, $J_2(t) = -J_L(-d,t)$, so that currents entering the terminals are taken as positive. 
The central quantity of interest in this work will be the zero frequency cross-correlated current fluctuation averaged over one period $T$ of the drive \cite{Jonckheere12}, namely
\begin{equation}
	\mc S_{23} = 2\int_0^T \frac{dt}{T} \int_{-\infty}^{+\infty} dt' \left[ \lan J_2(t) J_3(t') \ran - \lan J_2(t) \ran \lan J_3(t') \ran \right].
\end{equation}
Using the bosonization identity in equation \eqref{eq:bosonization}, it is possible to express $\mc S_{23}$ in terms of the bosonic Green's function $\mc G(\tau) = \lan \left[ \phi_{R/L}(0,\tau) - \phi_{R/L}(0,0)\right] \phi_{R/L}(0,0) \ran$ and the voltage signals only. It is worth noticing that $\mc G(\tau)$ does not depend on the spatial variable $x$ due to the assumption of local tunneling at the QPC.

After some algebra the only remaining contribution is found to be
\begin{equation}
\label{eq:S_23}
	\mc S_{23} = - 4 (e^*)^2 |\lambda|^2 \int_0^T \frac{dt}{T} \int_{-\infty}^{+\infty} dt' \cos \left\{ e^* \int_{t'}^{t} \left[ V_R(\tau) - V_L(\tau) \right] d \tau \right\} e^{2\nu \mc G(t'-t)}
\end{equation}
with $\lambda=\Lambda/(2 \pi a)$.

\subsection{Hanbury-Brown and Twiss setup}
\label{HBT}

In the context of EQO, Hanbury-Brown and Twiss (HBT) interferometry is performed by driving one of the input terminals with a periodic signal $V_{R/L}(t)$ while grounding the opposite one \cite{Hanbury56,Hanbury56b,Bocquillon12}.
To obtain the noise in this configuration, it is useful to expand the cosine in equation \eqref{eq:S_23} using the Fourier series representation of the periodic function $e^{-i \chi_{R/L}(t)}$, with $\chi_{R/L}(t) = \int_0^t [e^* V_{R/L}(t') - q_{R/L}\omega] dt'$. Physically, $\chi_{R/L}(t)$ is the phase accumulated by a Laughlin quasiparticle under the action of $V_{R/L}(t)$ and $q_{R/L}$ is the total charge injected during one period of the drive. The latter is linked to the DC component of the signal, namely \ $ \omega q_{R/L}  = e^* \int_0^T \frac{dt}{T} V_{R/L}(t)$ with $\omega= 2 \pi/T$ \cite{Rech17}.
Introducing the Fourier series $e^{-i \chi_{R/L}(t)} = \sum_l p_l(q_{R/L}) e^{-i l \omega t}$ and the Fourier transform $\hat P_{g}(E) = \int_{-\infty}^{+\infty} dt e^{i E t} e^{g \mc G(t)}$, with $g$ a generic real parameter, \cite{Braggio00} one gets
\begin{align}
\label{eq:S_HBT}
	\mc S_{R/L}^{\rm HBT}(q_{R/L})
	& = - 2(e^*)^2 |\lambda|^2 \sum_{l=-\infty}^{+\infty} \left|p_l(q_{R/L})\right|^2 \times \nonumber \\
	& \quad \times \left\{ \hat P_{2\nu} \left[ \left(q_{R/L} + l \right) \omega \right] + \hat P_{2\nu} \left[ - \left(q_{R/L} + l \right) \omega \right]\right\} .
\end{align}

We will focus on the interesting case of a periodic train of Lorentzian voltage pulses, whose Fourier coefficients are \cite{Dubois13,Rech17,Vannucci17,Grenier13} 
\begin{equation}
	p_l(q_{R/L}) = q_{R/L} \gamma^l \sum_{s=0}^\infty (-1)^{s} \frac{\Gamma(l+s+q_{R/L})}{\Gamma(1-s+q_{R/L})} \frac{\gamma^{2s}}{s! (s+l)!} .
\end{equation}
Here $\gamma=e^{-2\pi \eta}$ and $\eta$ is the ratio between the half width at half maximum of each Lorentzian peak and the period $T$ \cite{Glattli16,Glattli17,Glattli_Patent}.

\subsection{Hong-Ou-Mandel setup}
\label{HOM}

A solid-state equivalent of the photonic HOM experiment is reproduced when identical excitations collide at the QPC \cite{Dubois13b,Hong87,Jonckheere12,Jullien14,Bocquillon13}. Contacts 1 and 4 are thus driven with identical periodic voltage signals, the only difference being a tunable time delay $t_D$ between them. Here, we will consider a broader class of possible HOM collisions, where the left and right contact are driven with an identical shape in time but different amplitudes, namely $V_L(t) = \frac{q_L}{q_R} V_R(t+t_D)$. This choice allows us to elaborate on previous results regarding symmetric HOM interferometry, while also discussing some original results about generalized asymmetric collisions.

Following the same procedure we used for the HBT signal we immediately get
\begin{align}
\label{eq:S_HOM}
	\mc S^{\rm HOM}(q_R,q_L,t_D)
	& = - 2(e^*)^2 |\lambda|^2 \sum_{l=-\infty}^{+\infty} \left|\tilde p_l(q_R,q_L,t_D)\right|^2 \times \nonumber \\
	& \quad \times \left\{ \hat P_{2\nu} \left[(l + q_R - q_L)\omega \right] + \hat P_{2\nu} \left[-(l + q_R - q_L)\omega \right] \right\} ,
\end{align}
with coefficients $\tilde p_l(q_R,q_L,t_D) = \sum_m p_{l+m}(q_R) p^*_m(q_L) e^{i m \omega t_D}$.
It is customary to normalize the HOM noise with respect to the value expected for the random partitioning of a single source, namely the HBT signal \cite{Jonckheere12,Bocquillon13}. For this purpose we define the ratio
\begin{equation}
\label{eq:def_R}
	\mc R(q_R,q_L,t_D) = \frac{\mc S^{\rm HOM}(q_R,q_L,t_D) - \mc S^{(0)}}{\mc S_R^{\rm HBT} (q_{R})+ \mc S_L^{\rm HBT}(q_{L}) - 2\mc S^{(0)}}.
\end{equation}
Notice that we have subtracted the equilibrium noise $\mc S^{(0)} = - 4(e^*)^2 |\lambda|^2 \hat P_{2\nu}(0)$ both to the numerator and the denominator in equation \eqref{eq:def_R} \cite{Rech17,Bocquillon13}. 

Before getting into the details of the results, it is worth noticing that our HOM interferometer differs substantially from a gauge-transformed HBT setup.
Indeed, there is no gauge transformation that is able to map the equation of motion for the HOM configuration into an effective HBT with $V'_R(t) = V_R(t) - V_L(t)$ and $V'_L(t)=0$. However, the HOM noise in equation \eqref{eq:S_HOM} looks like the HBT contribution due to the peculiarity of point-like tunneling \cite{Ronetti18}.
This is linked to the fact that the tunneling amplitude for point-like tunneling between opposite edge states is energy independent in the Landauer-B\"uttiker picture of quantum transport. Any similarity between our HOM setup and an effective HBT would vanish for more complicated tunneling geometries such as multiple QPC or extended contacts \cite{Chevallier10,Dolcetto12}, where the transmission amplitude acquires a dependence upon energy \cite{Vannucci15,Ronetti16}.

\begin{figure}
	\centering
	\begin{overpic}
	[width=0.45\linewidth]{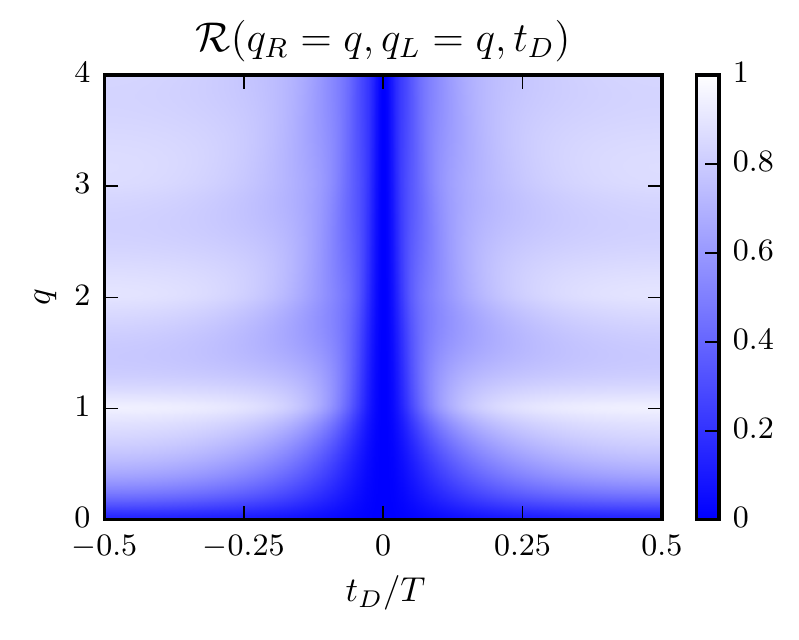}
	\put(1,70){a)}
	\end{overpic}
	\begin{overpic}
	[width=0.45\linewidth]{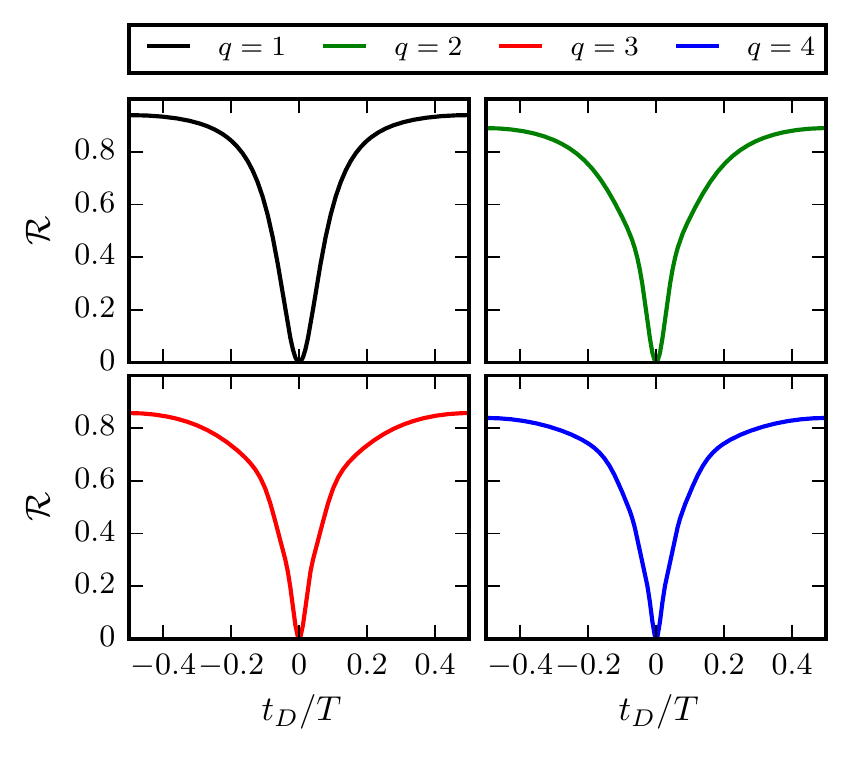}
	\put(1,70){b)}
	\end{overpic}
	\begin{overpic}
	[width=0.45\linewidth]{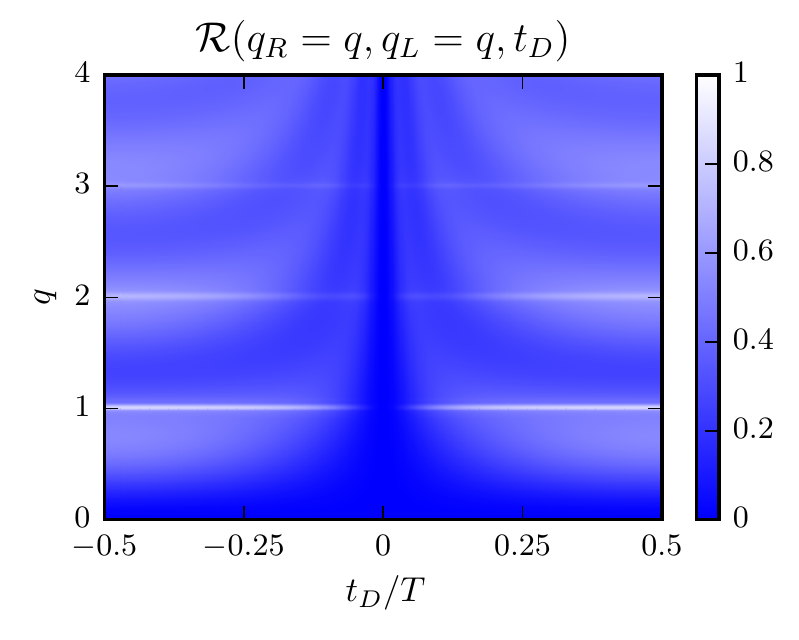}
	\put(1,70){c)}
	\end{overpic}
	\begin{overpic}
	[width=0.45\linewidth]{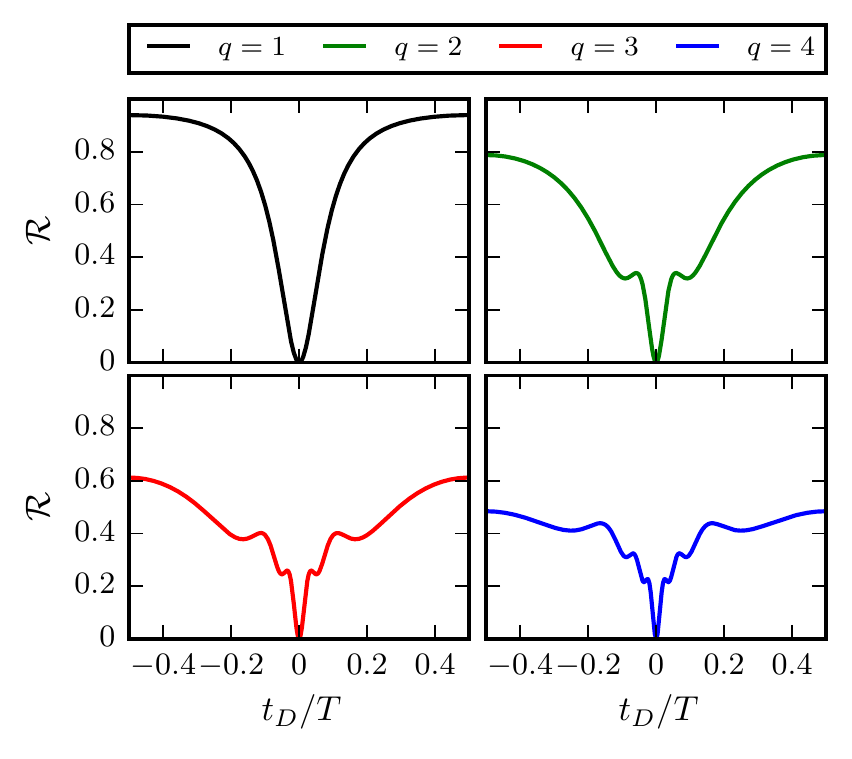}
	\put(1,70){d)}
	\end{overpic}
	\caption{HOM ratio $\mc R$ for collisions of identical excitations as a function of the delay $t_D$ (in units of the period $T$) and the number $q$ of injected charges. Panels a) and b) refer to the case of integer filling factor, while panels c) and d) correspond to fractional filling $\nu=1/3$. Curves in b) and d) are extracted respectively from the density plots in a) and c) by cutting at $q=1$, 2, 3 and 4. The temperature is $\theta=0.01 \, \omega$, the dimensionless width of the pulses is $\eta=0.04$ and the high energy cut-off is $v/a= 10 \, \omega$.}
	\label{fig:q_vs_q}
\end{figure}

\section{Results}
\label{Results}

\subsection{Symmetric HOM collisions}

We first look at the HOM ratio $\mc R$ in the case of integer filling factor ($\nu=1$) and identical drives ($q_R=q_L=q$). This case is reported in panels a) and b) of Figure\ \ref{fig:q_vs_q}. Here, exact results based on the scattering matrix formalism are available in the literature, both on the theoretical and experimental sides \cite{Dubois13,Dubois13b}. We observe that the noise is totally suppressed when colliding packets arrive simultaneously at the QPC ($t_D=0$). This is consistent with the anti-bunching effect expected for identical fermionic particles colliding at the QPC, which forces two identical fermions to leave the interferometer on opposite output arms, thus causing a drop in the fluctuations (the so-called Pauli dip)  \cite{Dubois13b,Jonckheere12,Bocquillon13}. However, we point out that a complete suppression of the noise is observed also for non integer values of $q$, i.e.\ when colliding packets are formed by several particle-hole pairs instead of a single electron-like excitation \cite{Levitov96,Keeling06}.
We also note that brighter regions in Figure\ \ref{fig:q_vs_q}a occur in correspondence of integer values of $q$. This is a signature of the robustness of Levitons at integer charge, namely a reduced contribution of particle-hole pairs. Finally, the qualitative behavior of $\mc R$ as a function of time delay is almost unaffected by variations of $q$ and shows a unique dip.

Things are dramatically different in the fractional regime, where new features linked to the strongly-correlated FQH phase come into play \cite{Ronetti17}. In Figure\ \ref{fig:q_vs_q}c we report the behavior of the HOM ratio for filling factor $\nu=1/3$. One still observes a complete dip in the ratio $\mc R$ when a null time delay is tuned between colliding packets, despite the presence of anyonic quasiparticles in the system. Once again, we ascribe this particular behavior to the geometry of our interferometer, which consists of a single QPC connecting the upper and lower edge. Such a tunneling geometry does not allow for closed-loop trajectories of one Laughlin quasiparticle around another, thus preventing any possible effect due to fractional statistics \cite{Chamon97}. At the same time, we note that the simple structure consisting of one single dip at $t_D=0$ is replaced by a much richer phenomenology. Indeed, Figure\ \ref{fig:q_vs_q}c shows the emergence of sub-dips for values $q>1$. This is even more evident in Figure\ \ref{fig:q_vs_q}d, where we have isolated the behavior of $\mc R$ at integer values of $q$. We note that the number of these sub-dips is given by $2q-2$. 

Such an unexpected behavior suggests that the injected wave-packet reorganizes into several sub-peaks at the output of the QPC. Indeed, due to the presence of a single edge channel, one cannot interpret this structure as a consequence of a spin-charge separation mechanism \cite{Wahl14,Freulon15,Marguerite16}. Moreover, let us notice that there is no coupling with bulk degrees of freedom in the model presented in Section \ref{Model}. The sub-dip structure cannot depend on the distance $d$ between the injector and the QPC as well (see Figure\ \ref{fig:setup_hom}), since such a parameter only manifests itself as an additional constant phase shift of the coefficients $p_l$ and $\tilde p_l$, which is clearly washed out in our final results (see equations \eqref{eq:S_HBT} and \eqref{eq:S_HOM}).
We can thus conclude that there is strong evidence of a rearrangement of the injected charge density outgoing from the QPC. This phenomenology can be explained in terms of a crystallization of $q$-Levitons occurring in the time domain, differently from what is typically discussed in condensed matter systems, and induced by the strong correlation characterizing the FQH state \cite{Ronetti17}.

Two additional features are worth noticing. First, the curves for $q=1$ at $\nu=1$ and $\nu=1/3$ (black lines in Figures\ \ref{fig:q_vs_q}b and \ref{fig:q_vs_q}d) are identical, which means that the Lorentzian peak carrying exactly one electron generates the same HOM ratio at both integer and fractional filling factor. Indeed, it was shown analytically in\ \cite{Rech17} that a remarkable factorization in the dependence on temperature and filling factor arises for this particular voltage drive, leading to a universal form of $\mc R(q_R=1, q_L=1, t_D)$ for each value of $\theta$ and $\nu$. This universality is consistent with the previously discussed picture, showing the impossibility for a single electronic wave-packet to crystallize. Second, the uniqueness of integer Lorentzian pulses is even more evident in the fractional regime when looking at the behavior for long delays between colliding packets (i.e.\ $t_D/T \approx \pm 0.5$). We observe that generic values of $q$ lead to a rather low HOM noise compared with the HBT contribution, which means that composite objects consisting of several particle-hole pairs are colliding at the beam splitter. However, integer Lorentzian pulses generate bright horizontal lines in Figure\ \ref{fig:q_vs_q}c. This is another striking confirmation of the fact that $q$-Levitons are minimal excitation states also in the FQH regime \cite{Rech17,Vannucci17}. Moreover, this is a signature of the fact that, differently from recently discussed periodic sources based on driven anti-dots \cite{Ferraro15}, a train of Lorentzian voltage pulse cannot be used as an on-demand source of fractionally charged quasiparticles \cite{Rech17}.

\begin{figure}
	\centering
	\begin{overpic}
	[width=0.45\linewidth]{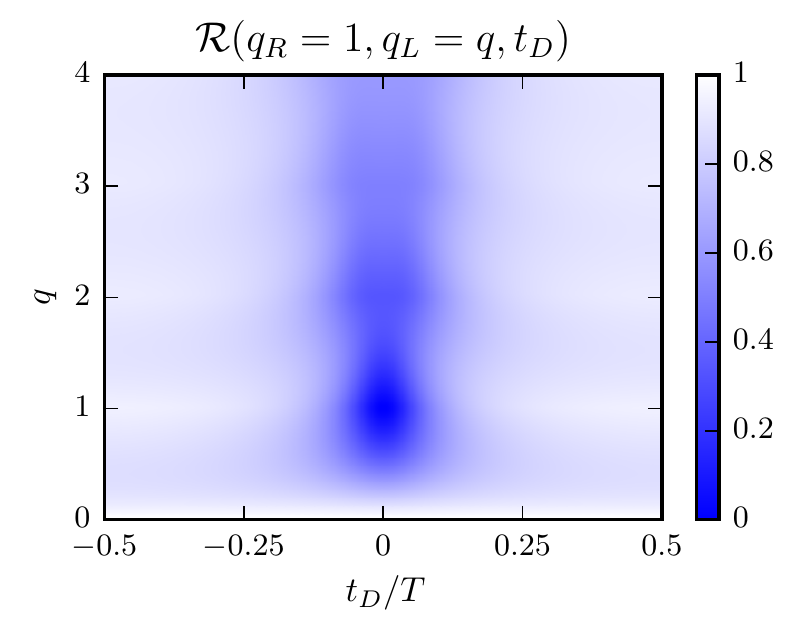}
	\put(1,70){a)}
	\end{overpic}
	\begin{overpic}
	[width=0.45\linewidth]{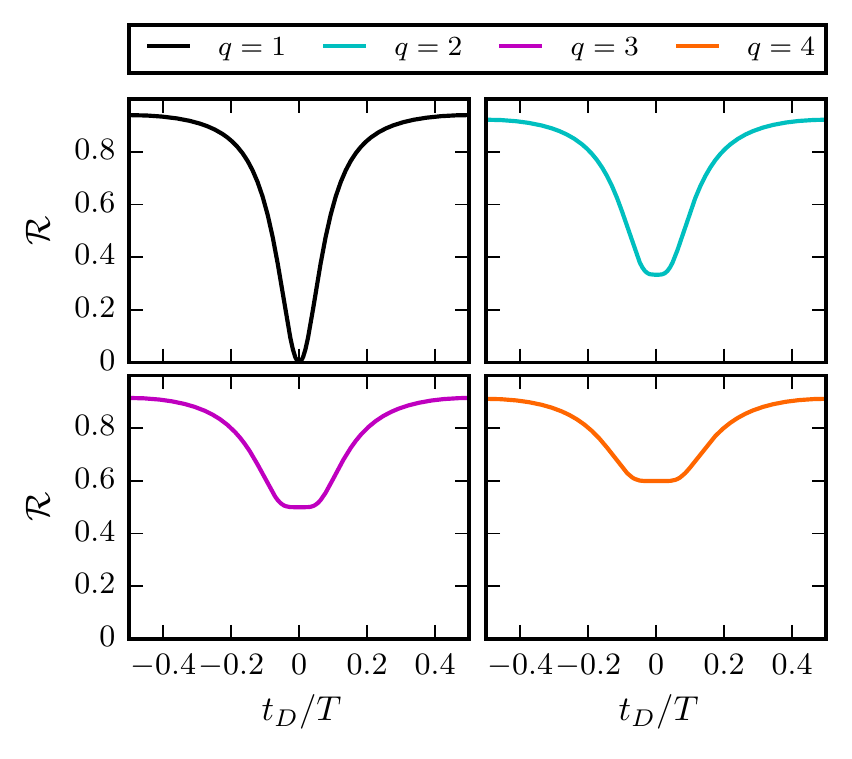}
	\put(1,70){b)}
	\end{overpic}
	\begin{overpic}
	[width=0.45\linewidth]{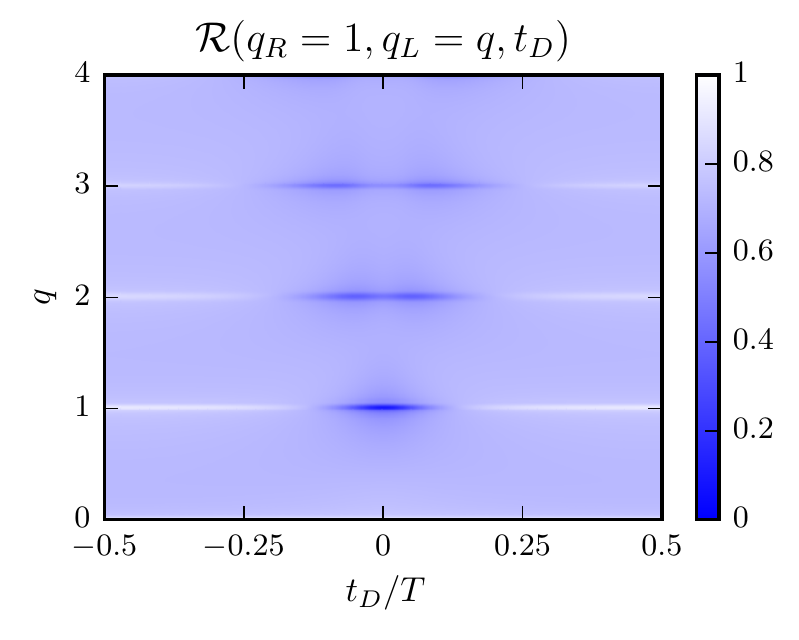}
	\put(1,70){c)}
	\end{overpic}
	\begin{overpic}
	[width=0.45\linewidth]{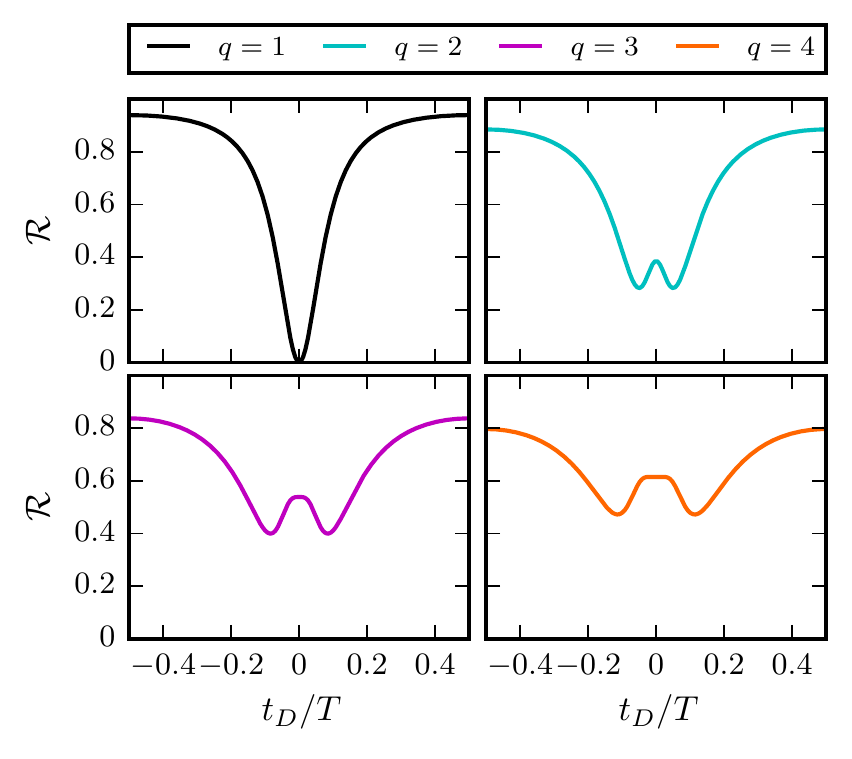}
	\put(1,70){d)}
	\end{overpic}
	\caption{HOM ratio $\mc R$ for collisions of integer ($q_R=1$) and generic ($q_L=q$) pulses as a function of the delay $t_D$ and the charge $q$ injected in the left-moving edge. Panels a) and b) refer to the case of integer filling factor, while panels c) and d) correspond to fractional filling $\nu=1/3$. Curves in b) and d) are extracted respectively from a) and c) at $q=1$, 2, 3 and 4. The temperature is $\theta=0.01 \, \omega$, the dimensionless width of the pulses is $\eta=0.04$ and the high energy cut-off is $v/a=10 \omega$.}
	\label{fig:1_vs_q}
\end{figure}

\subsection{Asymmetric collisions}

We now examine what happens when wave-packets carrying different charges are injected in terminals 1 and 4 and sent to the collider. For simplicity, we inject integer Lorentzian pulses with $q_R=1$ on the upper right-moving edge, while the opposite contact is driven with a generic Lorentzian drive with tunable $q_L$ and $t_D$. Starting from the $\nu=1$ case, we observe that the total suppression of HOM noise is achieved only for $q_L=q_R=1$, as expected. However, different values of $q_L$ still generate a partial dip in the noise whose value is given by
\begin{equation}
\mc R(q_R,q_L, t_D=0 ) = \frac{q_L-q_R}{q_L+q_R},
\end{equation}
as one can infer by looking at Figure\ \ref{fig:1_vs_q}a. For instance, by fixing the value of $q_L$ we obtain the four curves shown in Figure\ \ref{fig:1_vs_q}b, which clearly demonstrate that the anti-bunching effect is not complete - although still present - when excitations with different charge collide at the QPC.

In the FQH regime the values of the ratio $\mc R$ at zero delay is slightly modified, while two unexpected sub-dips appear (see Figures\ \ref{fig:1_vs_q}c and \ref{fig:1_vs_q}d). Despite being extremely well visible in the case of integer $q_L$, they are almost totally washed out when non-integer pulses are considered. This is again due to the fact that only an integer Lorentzian drive generates clean states in the conductor. Moreover, the distance between these two side-dips progressively increases by increasing $q_L$. As deduced previously from Figure \ref{fig:q_vs_q}, the $q_L$-Levitons, with $q_L>1$, are re-arranged in a crystallized structure, while the single Leviton wave-packet maintains a trivial pattern. Therefore, the latter is not suitable to resolve the more complicated crystal of $q_L$-Levitons in the type of experiment presented in Figure \ref{fig:1_vs_q}. As a consequence, the number of side-dips is totally unrelated to the specific value of $q_L$, in contrast to the configuration with identical drives. A better resolution would require the injection of a very narrow Leviton wave-packet from terminal $1$, ideally reaching the limit of a delta-like pulse: anyway, the generation of such a narrow pulse is beyond the state-of-art technology. Another possibility would be to increase the value of $q_R$, thus characterizing the crystal of $q_L$-Levitons with a comparably complicated wave-packet: in this case, a total number of $2 q_R$ side-dips would appear. 

\section{Wave-packet approach}
\label{wavepacket}

The current-current correlation generated in a collisional HOM experiment can be equally expressed in terms of the overlap between wavefunctions describing the Levitons impinging at the QPC. In this insightful formalism, a periodic train of $q$-Levitons is described by the following set of wavefunctions \cite{Glattli16}
\begin{equation}
\label{eq:phi}
\varphi_k(t)=\sqrt{\frac{\sinh\left(2\pi \eta\right)}{2}}\frac{\sin^{k-1}\left(\pi\frac{t}{T}-i \eta\right)}{\sin^{k}\left(\pi\frac{t}{T}+i \eta\right)}, \hspace{3mm} k=1,...,q .
\end{equation}
They generalize the set of single-electron wave functions introduced for the Lorentzian pulse \cite{Grenier13,Ferraro13,Moskalets15}, and form a complete orthonormal basis, thus satisfying the condition $\int_{0}^{T}\frac{dt}{T}\varphi_k(t)\varphi^{*}_{k'}(t)=\delta_{k,k'}$.
This set of wavefunctions is usually related to quantities called quasiparticle and quasihole correlation functions, defined as
\begin{align}
\label{eq:corr_def}
G^{(qp)}_{R/L}(t',t)& = \langle \Psi^\dagger_{R/L}(0,t')\Psi_{R/L}(0,t)\rangle=\frac{1}{2\pi a}e^{-i e^* \int_{{t'}}^{t} d\tau V_{R/L}(\tau)}e^{2\nu\mc G(t'-t)},\\
\label{eq:corr_def_qh}
G^{(qh)}_{R/L}(t',t)& = \langle \Psi_{R/L}(0,t')\Psi^\dagger_{R/L}(0,t)\rangle=\frac{1}{2\pi a}e^{i e^* \int_{{t'}}^{t} d\tau V_{R/L}(\tau)}e^{2\nu\mc G(t'-t)}.
\end{align}
We retained only temporal variables in equation\ \eqref{eq:corr_def}, since the space- and time-dependent field $\Psi_{R/L}(x,t)$ evolves only through the combination $x \mp vt$, due to the chirality of Laughlin edge states. Contrarily to ordinary optics, this correlator does not vanish even at equilibrium, due to the different nature of the ground state in fermionic systems. Since one is mainly interested in characterizing deviations from equilibrium introduced by the voltage drives, it is useful to define the excess quasiparticle correlators as
\begin{equation}
\label{eq:exc_corr_def}
\Delta G^{(qp)}_{R/L}(t',t)=G^{(qp)}_{R/L}(t',t)-G^{(0)}_{R/L}(t'-t),
\end{equation}
where $G^{(0)}_{R/L}(t'-t)=\frac{1}{2\pi a}e^{2\nu\mc G(t'-t)}$ corresponds to the correlator when $V_L=V_R=0$. Their expressions in terms of functions in equation \eqref{eq:phi} read
\begin{equation}
	\Delta G^{(qp)}_{R/L}(t',t) =-2 iG_0 (t' - t)  \sin \left( \frac{\pi(t' - t)}{T} \right) \sum\limits_{k=1}^{q_{R/L}}\varphi_{k,R}(t)\varphi^{*}_{k,L}{}(t'),
	\label{eq:DeltaG}
\end{equation}
where $\varphi_{k,R}(t)=\varphi_k(t)$ and $\varphi_{k,L}(t)=\varphi_{k}(t+t_D)$ and 
\begin{equation}
G_0(\tau) = \frac{1}{2\pi a} \left[ \frac{\pi k_{\rm B} \theta \tau}{\sinh\left( \pi k_{\rm B} \theta \tau \right) \left(1+i \frac{v \tau}{a}  \right)}\right]^{\nu}.
\label{G_0}
\end{equation}

It is instructive to note that, in the ordinary metallic case ($\nu=1$) and in the limit of infinite period, equation\ \eqref{eq:DeltaG} reduces to the so called single-electron coherence function, a crucial tool in the framework of electron quantum optics \cite{Grenier13,Ferraro13}. Excess correlators for quasiholes can be defined similarly as $\Delta G^{(qh)}_{R/L}(t'-t) = G^{(qh)}_{R/L}(t',t)-G_0(t'-t)$, yielding 
\begin{equation}
\label{eq:DeltaGqh}
\Delta G^{(qh)}_R = 2 i G_0(t'-t)\sin\left(\pi \frac{t'-t}{T}\right)\sum\limits_{k=1}^{q_{R/L}}\varphi^{*}_{k,R}(t)\varphi_{k,L}{}(t').
\end{equation}

By using equations \eqref{eq:corr_def} and the definitions of excess correlators for quasiparticles and quasiholes into Eqs. \eqref{eq:S_23} and \eqref{eq:def_R}, the ratio $\mathcal{R}$ can be expressed directly as (see Appendix \ref{AppA} for more details)

\begin{equation}
	\label{eq:ratio_SM}
	\mathcal{R}=1-\frac{\sum\limits_{k=1}^{q_R}\sum\limits_{k'=1}^{q_L}\sum\limits_{p=1}^{+\infty}\sum\limits_{p'=1}^{+\infty}\Re\left[w_{pp'}^{k}g_{k'p}(t_D)g^{*}_{k'p'}(t_D)\right]}{\frac{1}{2}\left(v_{q_R}+v_{q_L}\right)},
\end{equation}
with $g_{kp}(t_D)=\int_{0}^{T}\frac{dt}{T}\varphi_k(t+t_D)\varphi^{*}_p(t)$ and where the coefficients $w_{pp'}^{k}$ and $v_{q,r}$, which encode all the dependence on temperature and interaction, read
\begin{align}
	w_{pp'}^{k}&=\int_{0}^{T} \frac{dt}{T} \int\limits_{-\infty}^{+\infty}d\tau\hspace{1mm}\varphi_k(t)\varphi^{*}_k(t+\tau)\varphi_{p}(t)\varphi^{*}_{p'}(t+\tau)\sin^2\left(\frac{\pi\tau}{T}\right)G^2_0(\tau),\\
	v_{q_R}&=\sum\limits_{k=1}^{q_r}\int\limits_{-\infty}^{+\infty}d\tau \sin\left(\frac{\pi \tau}{T}\right)g^{*}_{kk}(\tau)G_0^2(\tau).\label{sup_eq:ratio_fin}
\end{align}

Equation \eqref{eq:ratio_SM} clearly shows the anti-bunching effect which arises when single-electron excitations as Levitons collide at a QPC. As the overlap between wavefunctions increases the value of $\mathcal{R}$ is reduced and eventually vanishes when the overlap between the two states is perfect.

Finally, starting from equation \eqref{eq:ratio_SM} one also gets a simplified formula for the integer case at zero temperature, which reads
\begin{equation}
	\label{eq:ratio_SM_integer}
	\mathcal{R}=1-\frac{2}{q_L+q_R}\sum\limits_{k=1}^{q_R}\sum\limits_{k'=1}^{q_L}\left|g_{k'k}(t_D)\right|^2,
\end{equation}
in accordance with previous results \cite{Jonckheere12,Glattli16}.

\section{Conclusions and perspectives}
\label{Conclusions}

In this work, we thoroughly reviewed the issue of HOM collisions in a FQH system. This can be done in a multi-terminal setup in presence of a QPC, which can be thought as an effective beam splitter for electron quantum optics experiments. By summarizing and further investigating previous results in the context of EQO, we have shown that synchronized collisions of identical excitations at the QPC always generate a total dip in the crossed shot noise, both at integer and fractional filling. However, the scenario for $\nu=1/3$ is enriched by the presence of several sub-dips in the HOM ratio, which are consistent with the crystallization of the wave-packet outgoing from the QPC. We have also explored a more generic HOM configuration involving Lorentzian pulses with different amplitudes, namely wave-packets carrying a different number of electrons on the opposite sides of the QPC. Here, the suppression of the shot noise is not complete, and two progressively separated side dips emerge in the FQH regime.

The generation and control of Levitons and, more in general, of single to few electronic excitations in one-dimensional ballistic conductors have attracted a huge interest in recent years and is actually at the forefront of research in condensed matter physics, and may lead to fascinating applications in quantum communication and quantum information.
However, several important issues, including for instance the role of decoherence \cite{Marguerite16,Cabart18} or screening \cite{Misiorny18} in the dynamics of single-electron excitations, are yet to be fully understood and will certainly deserve a lot of attention in years to come.

Further interesting developments could also involve the investigation of pseudorandom emission of Levitons, as recently proposed for the free fermion case \cite{Glattli18}. Besides closing the conceptual gap between time-periodic injection and individual wave-packet emission - Holy Grail of EQO \cite{Marguerite17} - this newly proposed emission protocol opens important perspectives for the FQH case, as it promises to strongly magnify the side dip features discussed in the present text.

As a final remark, let us notice that first experimental results about photo-assisted noise in the FQH regime are starting to be available \cite{Kapfer18}. This is of particular importance in a field like mesoscopic physics, where theory and experiments have always fed each other on the pursuit of new exciting discoveries. We believe that this is only the first step towards a full experimental EQO at fractional filling factor.

\section*{Acknowledgements}
We are grateful to E. Bocquillon, G. F\`eve and D. C. Glattli for useful discussions. L.V.\ acknowledges support from CNR SPIN through Seed project ``Electron quantum optics with quantized energy packets''. This work was granted access to the HPC resources of Aix-Marseille Universit\'e financed by the project Equip@Meso (Grant No. ANR-10-EQPX-29-01). It has been carried out in the framework of project ``1shot reloaded'' (Grant No. ANR-14-CE32-0017) and benefited from the support of the Labex ARCHIMEDE (Grant No. ANR-11-LABX-0033), all funded by the ``investissements d'avenir'' French Government program managed by the French National Research Agency (ANR). The project leading to this publication has received funding from Excellence Initiative of Aix-Marseille University - A*MIDEX, a French “Investissements d'Avenir” programme.

\section*{Authors contributions}
All the authors were involved in the preparation of the manuscript.\\
All the authors have read and approved the final manuscript.

\appendix 

\numberwithin{equation}{section}

\section{Calculation of the HOM ratio $\mathcal{R}$}
\label{AppA}
In this Appendix, we express the ratio defined in Eq. \eqref{eq:def_R} in terms of the set of wavefunctions given in Eq. \eqref{eq:phi}.\\ 
We can start by expressing the HOM current noise in terms of Green's functions in Eqs. \eqref{eq:corr_def} and \eqref{eq:corr_def_qh}  as \footnote{For notational convenience, from now on we have omitted the functional dependence of noise contributions introduced in the main text.}
\begin{equation}
	\label{sup_eq:noise1}
	\mc S^{\rm HOM}  = (\nu e)^2 |\Lambda|^2 \int_{0}^{T} \frac{dt}{T} \int\limits_{-\infty}^{+\infty} d \tau \left[G^{(qp)}_R(t+\tau,t)G^{(qh)}_L(t+\tau,t)+G^{(qh)}_R(t+\tau,t)G^{(qp)}_L(t+\tau,t)\right].
\end{equation}
It is interesting to recast the current noise directly in terms of the excess correlation functions $\Delta G_{R/L}^{(qp)}$ and $\Delta G_{R/L}^{(qh)}$, which encode all the information about Levitons. By using the relations $G_{R/L}^{(qp)}=\Delta G_{R/L}^{(qp)}+G_0$ and $G_{R/L}^{(qh)}=\Delta G_{R/L}^{(qh)}+G_0$ (we omitted here the time dependence for notational convenience), one finds
\begin{align}
	&\mc S^{\rm HOM}  =-\mc S^{(0)}+\mc S_{R}^{{\rm HBT}}+\mc S_{L}^{{\rm HBT}}\nonumber+\\&+ (\nu e)^2 |\Lambda|^2 \int_{0}^{T} \frac{dt}{T} \int\limits_{-\infty}^{+\infty} d \tau \left[\Delta G^{(qp)}_R(t+\tau,t)\Delta G^{(qh)}_L(t+\tau,t)+\Delta G^{(qh)}_R(t+\tau,t)\Delta G^{(qp)}_L(t+\tau,t)\right],\label{sup_eq:noise2}
\end{align}
where we have used the expressions of the equilibrium noise $\mc S^{(0)}$ and the HBT noise $\mc S^{{\rm HBT}}_{R,L}$ in terms of Green's functions, which read 
\begin{align}
	\label{sup_eq:noisevac}
	&\mc S^{(0)}  = 2 (\nu e)^2 |\Lambda|^2 \int\limits_{-\infty}^{+\infty} d \tau G^2_{0}(\tau),\\
	\label{sup_eq:noiseHBTR}
	&\mc S^{{\rm HBT}}_{R,L}  = (\nu e)^2 |\Lambda|^2 \int_{0}^{T} \frac{dt}{T} \int\limits_{-\infty}^{+\infty} d \tau \left[G^{(qp)}_{R/L}(t+\tau,t)+G^{(qh)}_{R/L}(t+\tau,t)\right]G_0(\tau).
\end{align}
By using the previous results the HOM ratio becomes
\begin{equation}
	\mathcal R = 1 - \frac{\mc N}{\mc D},
\end{equation}
with $\mc N$ and $\mc D$ given respectively by
\begin{align}
\label{eq:N}
	\mc N & = - \int_{0}^{T} \frac{dt}{T} \int\limits_{-\infty}^{+\infty} d \tau \left[\Delta G^{(qp)}_R(t+\tau,t)\Delta G^{(qh)}_L(t+\tau,t) + \right. \nonumber \\
	& \quad \left. + \Delta G^{(qh)}_R(t+\tau,t)\Delta G^{(qp)}_L(t+\tau,t)\right] , \\
\label{eq:D}
	\mc D & = \sum\limits_{r=R,L}\int_{0}^{T} \frac{dt}{T} \int_{-\infty}^{+\infty} d \tau \left[\Delta G^{(qp)}_r(t+\tau,t)+\Delta G^{(qh)}_r(t+\tau,t)\right]G_0(\tau).
\end{align}
By inserting equations \eqref{eq:exc_corr_def} and \eqref{eq:DeltaGqh} into the above equations, these two contributions can be further expressed in terms of the wavefunctions $\{\varphi_k\}$ as
\begin{align}
	\mc N & = \int_{0}^{T} \frac{dt}{T} \int\limits_{-\infty}^{+\infty} d \tau \left[\sum\limits_{k=1}^{q_R}\sum\limits_{k'=1}^{q_L}\varphi_k(t)\varphi^{*}_k(t+\tau)\varphi^{*}_{k'}(t+t_D)\varphi_{k'}(t+t_D+\tau)+\text{h.c.}\right] \times \nonumber \\
	& \quad \times \sin^2\left(\frac{\pi \tau}{T}\right)G^2_0(\tau) , \\
	\mc D & = \frac{1}{2}\sum\limits_{r=R,L}\int_{0}^{T} \frac{dt}{T} \int_{-\infty}^{+\infty} d \tau \left[i\sum\limits_{k=1}^{q_r}\varphi_k(t)\varphi^{*}_k(t+\tau)+\text{h.c.}\right]\sin\left(\frac{\pi \tau}{T}\right)G^2_0(\tau).
\end{align}
Finally, by introducing the overlap integrals of wavefunctions $g_{kp}(t_D)=\int_{0}^{T}\frac{dt}{T}\varphi_k(t+t_D)\varphi^{*}_p(t)$, the ratio acquires the form given in the main text
\begin{equation}
	\label{eq:ratio_SM_app}
	\mathcal{R}=1-\frac{\sum\limits_{k=1}^{q_R}\sum\limits_{k'=1}^{q_L}\sum\limits_{p=1}^{+\infty}\sum\limits_{p'=1}^{+\infty}\Re\left[w_{pp'}^{k}g_{k'p}(t_D)g^{*}_{k'p'}(t_D)\right]}{\frac{1}{2}\left(v_{q_R}+v_{q_L}\right)},
\end{equation}
where the coefficients are
\begin{align}
	w_{pp'}^{k}&=\int_{0}^{T} \frac{dt}{T} \int\limits_{-\infty}^{+\infty}d\tau\hspace{1mm}\varphi_k(t)\varphi^{*}_k(t+\tau)\varphi_{p}(t)\varphi^{*}_{p'}(t+\tau)\sin^2\left(\frac{\pi\tau}{T}\right)G^2_0(\tau),\\
	v_{q_r}&=\sum\limits_{k=1}^{q_r}\int\limits_{-\infty}^{+\infty}d\tau \sin\left(\frac{\pi \tau}{T}\right)g^{*}_{kk}(\tau)G_0^2(\tau).\label{sup_eq:ratio_fin_app}
\end{align}

\end{document}